# Role of the gravitomagnetic field in accelerating accretion disk matter to polar jets

J. Poirier* and G. J. Mathews**

*Center for Astrophysics, Department of Physics, University of Notre Dame, Notre Dame, IN 46556 USA*

We show that the motion of the neutral masses in an accretion disk orbiting a black hole creates a general-relativistic magnetic-like (gravitomagnetic) field that vertically accelerates neutral particles near the accretion disk upward and then inward toward the axis of the accretion disk. Even though this gravitomagnetic field alone does not achieve collimated jets, it is a novel means to identify one general relativistic effect from a much more complicated problem. In addition, as the accelerated material above or below the accretion disk nears the axis with a nearly vertical direction, a frame-dragging effect twists the trajectories around the axis thus contributing to the formation of collimated bipolar jets.

## I. INTRODUCTION

Although the existence of well collimated jets carrying large amounts of energy away from accreting black holes is well established observationally, we are still a long way from understanding how such well collimated jets are launched and stabilized (see review of 2014 [1]). Indeed, since the 1977 pioneering work of Blandford and Zanek (BZ) [2], it is generally believed that such jets are launched electromagnetically (EM) when a rotating black hole and/or accretion disk is threaded by magnetic field lines (the BZ effect). Indeed, there has been considerable progress in the development of fully general relativistic MHD simulations of black hole accretion disks and jets in two and three spatial dimensions [3-14]. As a complement to these numerical studies, our goal here is to present a physical, semi-analytic analysis of a simple general relativistic effect that may be difficult to identify in a fully general relativistic large-scale numerical simulation.

The purpose of this letter is to demonstrate a novel, purely general relativistic, mechanism (GEM) that contributes to one aspect of the formation and collimation of jets from black holes in addition to the electromagnetic (EM) BZ effect [2] noted above. In particular, we note that, although in a realistic simulation the effects of a magnetic field and the spin of the central black hole might dominate over the effect noted here, our intention is to point out that the GEM equations give a new means to understand this aspect of the purely general relativistic effect that occurs even in the absence of magnetic fields.

This complicated general relativistic effect is particularly easy to isolate and understand when one employs the linearized 1$^{st}$ post-Newtonian (1PN) approximation to the Einstein equations written in this 1PN form analogous to the equations of electromagnetism (EM); namely, gravitoelectromagnetism (GEM). Here, we discuss how the general relativistic Einstein equations in this 1PN approximation produce gravitoelectric and gravitomagnetic fields $\mathbf{E_g}$ and $\mathbf{B_g}$ from the moving masses of the accretion disk. In particular, the effects of the gravitomagnetic field $\mathbf{B_g}$ causes uncharged nearby moving masses with velocities similar to the accretion disk to experience a gravito-Lorentz-like force, $\mathbf{F}_g$. This general relativistic force accelerates moving neutral masses thus contributing to the formation of relativistic jet outflows even in the absence of electro-magnetic fields (EM). To our knowledge this GEM contribution to relativistic jet formation has not previously been noted.

Here, we show that the gravito-force can be large near a black hole accretion disk since this force is proportional to the large mass of the disk and depends quadratically on their high velocities. We analyze the magnitude and direction of this new and novel additional contribution to the formation of black hole jets.

## II. GRAVITOELECTROMAGNETIC (GEM) EQUATIONS

The linearized 1$^{st}$ post-Newtonian approximation (1PN) to general relativity has been widely employed in astrophysics. Moreover, it has been known for many years that these equations can be recast in Maxwell-like form [15-20]. The present work relies on the equations of GEM as deduced and summarized in Kaplan, Nichols, and Thorne (KNT [20]) in which the treatment of Ref. [16] was revised to include the harmonic gauge and the analogies with Maxwell's equations of EM were well specified. However, whereas KNT were primarily concerned with the evolution of merging binaries, the application here is to a simpler problem, i.e., the gravitomagnetic field due to a black hole accretion disk. We note, however, that this 1PN analysis should be taken with caution since it does not capture the Kerr spacetime. Nevertheless this analysis provides a transparent way to identify one general relativistic effect in an otherwise very complicated space-time. Although it is not exact, it illustrates an interesting general relativistic effect of non-negligible strength as we now show.

The spatial region of interest to this letter ($R > 9GM_{BH}/c^2$) is somewhat beyond the last stable circular orbit and well outside the black-hole horizon ($R_s = 2GM_{BH}/c^2$), and is associated with only moderately relativistic velocities ($(v/c)^2 \sim 0.1$). Thus 1PN is an adequate approximation for qualitatively analyzing the general relativistic accretion disk physics of interest here.

For a stationary and pressureless accretion disk, the diagonal components of the energy-momentum tensor in this limit become $T^{00} \approx \rho_g$, and $T^{jj} \approx 0$, where $\rho_g$ is an effective mass-energy density. Similarly, the space-time components of the energy-momentum tensor $T^{0i}$ can be represented as an effective gravito-mass-current density $\mathbf{J}_g = \rho_g \mathbf{v}$. The general relativistic GEM equations of KNT [20] then reduce to:

$$\nabla \cdot \mathbf{E}_g = -4\pi G \rho_g, \qquad (1)$$

$$\nabla \cdot \mathbf{B}_g = 0, \qquad (2)$$

$$\nabla \times \mathbf{E}_g = -\frac{\partial \mathbf{B}_g}{\partial t}, \qquad (3)$$

$$\nabla \times \mathbf{B}_g = 4\left(\frac{-4\pi G}{c^2}\mathbf{J}_g + \frac{\partial \mathbf{E}_g}{c^2 \partial t}\right), \qquad (4)$$

with G and c included to restore the physical constants. Equations (1-4) are obviously analogous to their electromagnetic (EM) Maxwellian counterparts.

The force on a test particle of mass *m* in the GEM fields $\mathbf{E}_g$ and $\mathbf{B}_g$ takes the form of a GEM Lorentz-like force, $\mathbf{F}_g$:
$$\mathbf{F}_g = m(\mathbf{E}_g + \mathbf{v} \times \mathbf{B}_g). \qquad (5)$$
This gravito-force $\mathbf{F}_g$ arises when a nearby uncharged body of mass *m* moves with velocity **v** in the gravitoelectromagnetic $\mathbf{E}_g$ and $\mathbf{B}_g$ fields produced by the accretion disk. Focus on the effects of $\mathbf{B}_g$ in Eq. (5): the direction of $\mathbf{F}_g$ is always perpendicular to **v** and thus the resulting acceleration from the $\mathbf{B}_g$ field changes only the direction of **v** and not its magnitude.

There are, however, three noteworthy differences between these GEM equations and Maxwell's equations in electromagnetic theory (EM):
  A. In Eq. (1) there is a minus sign which gives an attraction between two masses compared to a repulsive force for like charges in EM.
  B. In Eq. (4) there is a minus sign on the $\mathbf{J}_g$ term that gives repulsion between two masses with parallel velocities compared to EM theory which gives an attraction between two similarly moving, like charges.
  C. In Eq. (4) there is an extra factor of 4 compared with Eq. (1). Thus, for v near c, the gravitomagnetic repulsion of these two co-moving masses is four times greater than the gravitoelectric (gravitational) attraction; for smaller velocities, this ratio of four diminishes as $(v/c)^2$.

In the remainder of this paper we focus solely on the effects of the accretion disk (except for a brief mention of the central black hole at R=0 in the summary section). Our simple example is intended to show the strength and the effect of the general relativistic gravitomagnetic field arising from the accretion disk. We do this by utilizing the above 1PN approximation; namely, the first term in Eq. (4).

### III. GRAVITOMAGNETIC FIELD, $B_g$, OF AN ACCRETION DISK

For illustration, a non-evolving accretion disk is assumed so that $\mathbf{E}_g$ is constant and the value of $\mathbf{B}_g$ is calculated only from the current density $\mathbf{J}_g$ term in Eq. (4) arising from the moving masses of the accretion disk. The gravitomagnetic field, $\mathbf{B}_g$, due to the accretion disk is then calculated by numerically integrating the equivalent Biot-Savart law in GEM,
$$d\mathbf{B}_g = \left(\frac{-4G}{c^2}\right)\frac{I_g d\mathbf{l} \times \mathbf{r}}{r^3}, \qquad (6)$$
where $I_g$ is the gravitocurrent given by $I_g = \mathbf{J}_g \cdot \mathbf{A}$; the unit for $I_g$ is kg sec$^{-1}$ and $\mathbf{B}_g$ is sec$^{-1}$.

As an example, we consider an accretion disk of one solar mass around a central black hole of ten solar masses with a corresponding Schwarzschild radius of 30 km. We note that this is a somewhat large disk mass ratio and is only appropriate for a somewhat low mass black hole. Nevertheless, a disk mass this large has been produced, for example, in simulations of jet formation in collapsars (e.g. Ref. [6] and references therein). The coordinate system is chosen as in Fig. 2a with an accretion disk located at midplane (z = 0) and rotating in a counterclockwise direction as seen from above with a cylindrical radius $R = (x^2 + y^2)^{1/2}$. Consider an inner accretion disk extending from 1.0 ($R_1$ = 135 km, slightly outside the innermost stable circular orbit) to 2.0 ($R_2$ = 270 km) with Keplerian orbital velocities v = 0.33c at $R_1$ that vary as the inverse square root of the flat-space coordinate distance from the black hole. The accretion disk material is assumed to be uniformly distributed between $R_1$ and $R_2$. These conditions are consistent with the often-employed alpha disk model [21] for black hole accretion disks.

The resulting gravitomagnetic field lines of **$B_g$** are given in Fig. 1 with the direction of the field denoted by arrows (note that this direction is opposite to that of **B** in an EM problem of positive charges moving in the same direction). At R = 0 and z = 0 the resulting magnitude of $B_{gz}$ = 23 sec$^{-1}$, which is independent of the test particle's mass. Taking this value as a lower limit for **$B_g$** above and near the accretion disk, for a test particle of velocity 0.33c, the vertical acceleration is > 23 x 10$^8$ m/sec$^2$, independent of the test particle's mass. The result for **$B_g$** at the origin can be scaled: e.g., if the ratio of $M_{BH}/R_1$ remains as in the above example, then $v_{1x}$ is unchanged and $B_{gz}$ scales as $M_{disk}/R_1$. Table I gives scaled examples of $B_{gz}$ for higher and lower values of masses and accretion disk radii. This extrapolation is not valid when applied to supermassive black holes since they have much lower ratios of $M_{disk}$ to $M_{BH}$.

Table I

| $M_{BH}$ ($M_{sun}$) | $M_{disk}$ ($M_{sun}$) | $R_1$(km) | $B_{gz}$ (sec$^{-1}$) |
|---|---|---|---|
| 3 | 0.2 | 41 | 15 |
| **10** | **1** | **135** | **23** |
| 30 | 4 | 405 | 31 |

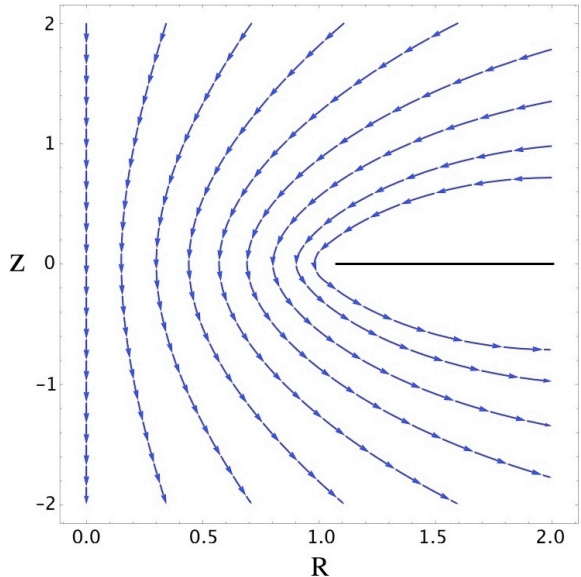

FIG. 1. Gravitomagnetic field showing field lines of **$B_g$** created by accretion disk material rotating counter-clockwise as seen from above with a radius R from 135 km (R = 1.0) to 270 km (R = 2.0). Other scaled possibilities yielding the same field line shape are given in Table I with the resulting magnitudes for $B_{gz}$ at the origin

## IV. PARTICLE TRAJECTORIES

Figs. 2(a), (b), (c) show numerically integrated typical trajectories calculated for test masses accelerated in the gravitomagnetic field $\mathbf{B}_g$ which produces the force $\mathbf{F}_g$ from Eq. (5). These nearby test masses move in the $\mathbf{B}_g$ field of Fig. 1 with an initial velocity in the -x (counterclockwise) direction similar to the velocities of the moving masses in the accretion disk. It has been previously noted in point C of Section II that, for v/c near unity, the effects of $\mathbf{B}_g$ are four times greater than those of $\mathbf{E}_g$ from the disk.

The five trajectories in Figs. 2(a), (b), (c) correspond to initial locations (in units of $R_1$) of: x = 0; y = 1.1, 1.2, 1.3, 1.4, 1.5; with z = ± 0.2 (slightly above or below the accretion disk). The acceleration on these nearby masses is independent of their mass since $\mathbf{F}_g$ in Eq. (5) is proportional to their dynamical mass, while their acceleration is inversely proportional to their inertial mass, and the ratio of these two masses is unity. The trajectories were obtained by numerically integrating this acceleration.

Note in particular that near the accretion disk this gravitomagnetic $\mathbf{F}_g$ force is almost vertical, but as particles move upward (downward), away from the disk, the force becomes increasingly directed toward the +z (-z) axis helping to induce a collimation of the jet.

A "three-dimensional" representation of these trajectories is shown in Fig. 2(a); their projections onto two orthogonal planes are shown in Figs. 2(b) and 2(c). Trajectories with initial values of y > 1.5 are not shown since they were no longer directed inward toward the polar axis, and hence would not contribute to the collimated jet.

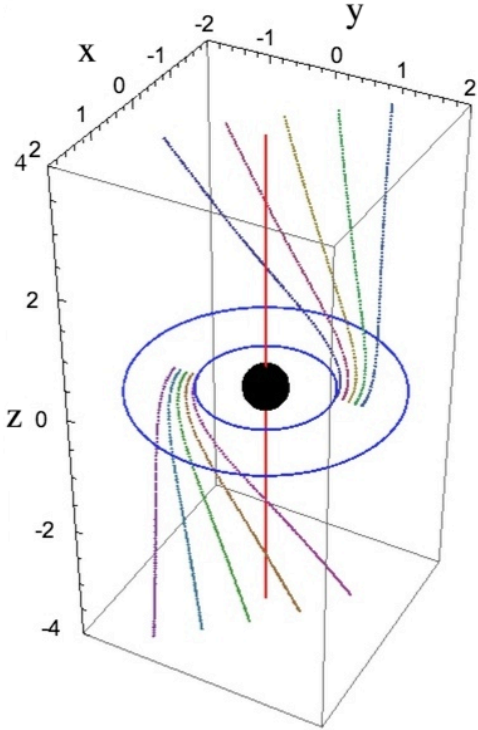

FIG. 2(a). A "three-dimensional" view of the resulting trajectories due to the GEM force from the accretion disk.

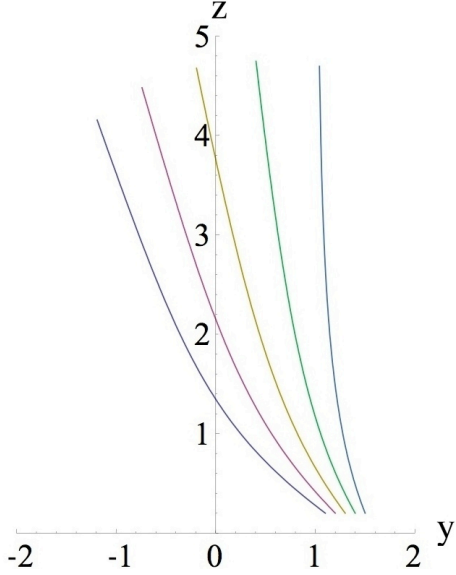 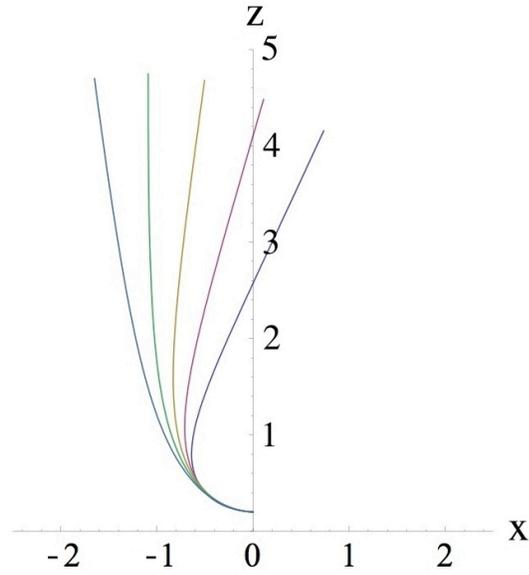

FIG. 2(b). The projection of the trajectories onto the yz-plane at x = 0.

FIG. 2(c). The projection onto the xz-plane at y = 0.

## V. CONTRIBUTION TO JET COLLIMATION ALONG THE POLAR AXIS

The particle trajectories calculated here are nearly vertical as they pass close to the z-axis. However, our calculation of the effect of the gravitomagnetic field of the accretion disk does not achieve well-collimated jets by itself. A number of other effects contribute to the formation of collimated jets including the BZ effect noted above. Also, most black holes have spin [22] so, for example, one should include the gravitomagnetic effects of the spinning black-hole and the Lense-Thirring frame-dragging effect due to the angular momentum of the accretion disk and a spinning central black hole that together would twist the trajectories near the polar axis to even smaller angles relative to the polar direction. A simulation of these added effects along with the GEM effects described here will be explored in a future paper. Our main point in this paper however is that, although the formation of relativistic jets is a complicated problem of many facets, we identify a novel means to easily clarify the role of one general relativistic effect.

The source of this gravitomagnetic field is from the moving masses of the disk and thus can be calculated in magnitude and direction from observable quantities. The resulting gravitomagnetic force in this astrophysical setting can be significant. For example, the time it takes to accelerate a test particle via this effect from near the plane of the accretion disk to the top of Fig. 2(a) (a height of >500 km) and to change its direction by ~90 degrees is independent of the test particle's mass and takes less than 0.01 seconds.

ACKNOWLEDGMENTS

We wish to acknowledge Profs. Chris Kolda, Samir Bose, and Bill McGlinn for assistance with the post-Newtonian formulation; Profs. Umesh Garg, Peter Garnavich, and John LoSecco for helpful suggestions; Jared Johnson, Adrien Saremii, and Aaron Sawyer for their numerical calculations; Prof. Mark Caprio, Tom Catanach, and Christina Horr for help with the figures. This work has been supported in part by the Mt. CubaAstronomical Foundation, and by the U.S. Department of Energy under DOE grant DE-FG02-95-ER40934.

*poirier@nd.edu
**gmathews@nd.edu